\documentclass[aps,prl,twocolumn,groupedaddress]{revtex4}
\pdfoutput=1
\usepackage{graphicx}
\usepackage{subfigure}
\usepackage{amssymb}
\usepackage{amsmath}
\usepackage{mathrsfs}

\newcommand{\mbf}[1]{\boldsymbol{#1}}

\newcommand{\Rey}{\textit{Re}}
\newcommand{\Wi}{\textit{Wi}}

\makeatletter

\newcommand{\numtoRoman}[1]{\expandafter\@slowromancap\romannumeral #1@}
\makeatother

\begin{document}

%\preprint{}

\title{Active and hibernating turbulence in minimal channel flow of Newtonian and polymeric fluids}
%\title{Active and hibernating turbulence in channel flow of Newtonian and polymeric fluids}


\author{Li Xi}
\author{Michael D. Graham}
\email[Corresponding author: ]{graham@engr.wisc.edu}
%\thanks{}
\affiliation{Department of Chemical and Biological Engineering, University of Wisconsin-Madison,  WI 53706-1691}


\date{\today}

\begin{abstract}
%ABSTRACT NEEDS TO BE SHORTER

%\input{abstract}
% !TEX root = LiXi_hibernation.tex
%
Turbulent channel flow of drag-reducing polymer solutions is simulated in minimal flow geometries. Even in the Newtonian limit, we find intervals of ``hibernating'' turbulence that display many features of the universal maximum drag reduction (MDR) asymptote observed in polymer solutions: weak streamwise vortices, nearly nonexistent streamwise variations and a mean velocity gradient that quantitatively matches experiments. As viscoelasticity increases, the frequency of these intervals also increases, while the intervals themselves are unchanged, leading to flows that increasingly resemble MDR.
\end{abstract}

%\keywords{}
\maketitle

% leave a empty line between sections to ensure change of paragraph.


% !TEX root = LiXi_hibernation.tex
%
The energy dissipated in turbulent channel or pipe flow of a liquid can be
dramatically reduced by low levels of long-chain polymer additives~\citep{Virk_AICHEJ1975, Graham_ReologyReviews2004, White_Mungal_ARFM2008}. The most striking qualitative feature of this phenomenon is the existence of a so-called maximum drag reduction (MDR) asymptote~\citep{Virk_AICHEJ1975}.  For a given flow geometry at a given pressure drop (i.e.~at a given Reynolds number $\Rey$), there is an asymptotic maximum flow rate that can be achieved through addition of polymers. Changing the concentration, molecular weight or even the chemical structure of the additives has no effect on this asymptotic value. 
 This universality is the major puzzle of drag reduction.

Turbulent flow in or near the MDR regime displays important differences from Newtonian turbulence.
% or turbulence at low levels of drag reduction (LDR). 
% The MDR asymptote is not a high Reynolds number phenomenon  -- it is experimentally observed at values of $\Rey$ all the way down into the laminar-turbulent transition regime \cite{Virk_AICHEJ1975}. 
 Its most commonly discussed signature is a distinctive mean velocity profile $U_\mathrm{mean}(y)$ that displays clear log-law behavior well-approximated by a formula given by Virk: $U_\mathrm{mean}^+ = 11.7\:\ln y^+ - 17.0$~\citep{Virk_AICHEJ1975} (Superscript ``+'' denotes quantities nondimensionalized in inner velocity and length scales $\sqrt{\tau_\mathrm{w}/\rho}$ and $\eta /\sqrt{\rho\tau_{w}}$; $\tau_\mathrm{w}$ is the time- and area averaged wall shear stress, $\eta$ and $\rho$ are fluid viscosity and density and $y$ is distance from the wall.) Additionally, streamwise vortices, which dominate near-wall dynamics in Newtonian  turbulence, are significantly weakened at MDR. Low-speed streaks become much less wavy in the streamwise direction and the streak spacing is substantially larger~\citep{White_Mungal_EXPFL2004,Housiadas_Beris_POF2005,Li_Sureshkumar_JNNFM2006,Xi_Graham_submitted2009}. 
%As MDR is approached, the slope of the mean velocity profile is increased across the entire channel, and approaches the limit of $11.7$ at MDR~\citep{Warholic_Hanratty_EXPFL1999, Ptasinski_Nieuwstadt_JFM2003, Min_Choi_JFM2003b, Dubief_Lele_JFM2004, Li_Sureshkumar_JNNFM2006, Xi_Graham_submitted2009}. 
In addition, the Reynolds shear stress near MDR is substantially smaller %, by an order of magnitude or more in some cases, 
than in Newtonian turbulence~\citep{HulsenMDR01,Warholic_Hanratty_EXPFL1999,Warholic01,Ptasinski_Nieuwstadt_JFM2003}. Indeed, the smallness of Reynolds shear stress is a central issue in a recent phenomenological model of MDR \cite{Procaccia_Lvov_RMP2008}.

Based on these observations, many researchers have suggested that turbulence in the MDR regime is ``transitional'' \cite{White_Mungal_ARFM2008} or ``marginal'' \cite{Procaccia_Lvov_RMP2008} in some sense that is not yet well-defined, and that the spatiotemporal flow structures that sustain turbulence in this regime are substantially different from those of normal Newtonian turbulence.  
%These observations indicate that the flow structures that dominate the MDR regime are substantially different from those of Newtonian turbulence. 
In the latter case, one simulation approach that has been fruitful in identifying the self-sustaining structures is the so-called minimal flow unit (MFU) approach~\cite{Jimenez_Moin_JFM1991,Webber97}. This approach identifies the smallest flow domain (at a given Reynolds number) in which turbulence can be sustained. Accordingly, temporally intermittent phenomena can be identified more readily than in a large box, where spatial averages incorporate different regions in which the instantaneous behavior may be very different.  
%Only one prior work \cite{Xi_Graham_submitted2009} has addressed dynamics in MFUs for viscoelastic flows, and 
The present work is the first to address the relevance of this approach for understanding MDR.

% TWS

%Also universal is the shape of the mean velocity profile as a function of distance from the channel wall. 

% !TEX root = LiXi_hibernation.tex
%

We focus on plane Poiseuille flow with fixed pressure drop. Streamwise, wall-normal and spanwise directions are denoted $x$, $y$, and $z$,
          respectively. The no-slip boundary condition applies at
           $y=\pm 1$ and periodic boundary conditions apply in
          $x$ and $z$; the periods in these directions are
          $L_{x}$ and $L_{z}$. Lengths are
          scaled with half-channel height $l$ and velocities with the Newtonian laminar
          centerline velocity $U$ at the given pressure drop. Time $t$
          is scaled with $l/U$ and pressure $p$ with $\rho U^2$. The governing equations are:
%\begin{eqnarray}%
%	\label{Eq_ns_momentum}%
%		\frac{\partial \mbf{v}}{\partial t} +
%		\mbf{v} \cdot \mbf{\nabla v} = -
%		\mbf{\nabla}p + \frac{\beta}{\Rey} \nabla^{2}\mbf{v} +
%		\frac{2\left(1 -\beta\right)}{\Rey \Wi}\left(\mbf{\nabla} \cdot
%		\mbf{\tau}_{\mathrm{p}}\right),%
%	\\%
%	\label{Eq_ns_continuity}%
%		\mbf{\nabla} \cdot \mbf{v} = 0,%
%	\\
%	\label{Eq_fenep_conformation}
%		\begin{split}
%			\frac{\Wi}{2} \left(
%			\frac{\partial \mbf{\alpha}}{\partial t} +
%			\mbf{v} \cdot \mbf{\nabla \alpha} -
%			\mbf{\alpha} \cdot \mbf{\nabla v} - \left( \mbf{\alpha} \cdot \mbf{\nabla v}
%			\right)^{\mathrm{T}} \right)
%		\\
%			+ \frac{\mbf{\alpha}}{1 - \frac{\mathrm{tr}(\mbf{\alpha})}{b}}
%			= \left( \frac{b}{b + 2} \right)
%			\mbf{\delta},%
%		\end{split}
%	\\%
%	\label{Eq_fenep_stress}%
%		\mbf{\tau}_{\mathrm{p}} = \frac{b + 5}{b} \left( \frac{\mbf{\alpha}}{1 -
%		\frac{\mathrm{tr}(\mbf{\alpha})}{b}} -\left( 1 - \frac{2}{b + 2} \right) \mbf{\delta}
%		\right).%
%\end{eqnarray}%
\begin{equation}%
	\label{Eq_ns_momentum}%
%		\frac{\partial \mbf{v}}{\partial t} +
%		\mbf{v} \cdot \mbf{\nabla v}
		\frac{D\mbf{v}}{Dt} = -
		\mbf{\nabla}p + \frac{\beta}{\Rey} \nabla^{2}\mbf{v} +
		\frac{2\left(1 -\beta\right)}{\Rey \Wi}\left(\mbf{\nabla} \cdot
		\mbf{\tau}_{\mathrm{p}}\right),
		\mbf{\nabla} \cdot \mbf{v} = 0%
\end{equation}
%\begin{equation}
%	\label{Eq_ns_continuity}%
%		\mbf{\nabla} \cdot \mbf{v} = 0,%
%\end{equation}
\begin{equation}
	\label{Eq_fenep_conformation}
		\begin{split}
			\frac{\Wi}{2} \left(
			\frac{D\mbf{\alpha}}{D t} 
%			+
%			\mbf{v} \cdot \mbf{\nabla \alpha}
			 -
			\mbf{\alpha} \cdot \mbf{\nabla v} - \left( \mbf{\alpha} \cdot \mbf{\nabla v}
			\right)^{\mathrm{T}} \right)=-\frac{b}{b+5}\mbf{\tau}_{\mathrm{p}},
%		\\
%			+ \frac{\mbf{\alpha}}{1 - \frac{\mathrm{tr}(\mbf{\alpha})}{b}}
%			= \left( \frac{b}{b + 2} \right)
%			\mbf{\delta},%
		\end{split}
\end{equation}
\begin{equation}			
	\label{Eq_fenep_stress}%
		\mbf{\tau}_{\mathrm{p}} = \frac{b + 5}{b} \left( \frac{\mbf{\alpha}}{1 -
		\frac{\mathrm{tr}(\mbf{\alpha})}{b}} -\left( 1 - \frac{2}{b + 2} \right) \mbf{\delta}
		\right).%
\end{equation}%
Eq.~(\ref{Eq_ns_momentum}) describes conservation of momentum and mass, and the polymer conformation and stress tensors $\mbf{\alpha}$ and $\mbf{\tau}_{\mathrm{p}}$ are described by the FENE-P constitutive equation for a bead-spring dumbbell model of a polymer in solution ((Eqs.~\ref{Eq_fenep_conformation}) and~(\ref{Eq_fenep_stress}))~\citep{Bird_Curtis_1987}. Reynolds number  $\Rey \equiv \rho U l / \eta$, Weissenberg number $\Wi \equiv 2 \lambda U/l$ ($\lambda$ is the polymer relaxation time), viscosity ratio $\beta \equiv \eta_\mathrm{s} / \eta$ ($\eta_{s}$ is the solvent contribution to zero-shear viscosity), and $b$ is the upper limit of the extension of polymer. All simulations reported here are at $\Rey=3600, \beta=0.97, b=5000$. The numerical integration procedure is described in \citep{Xi_Graham_submitted2009}.
%SAY SOMETHING ABOUT RESULTS AT DOUBLE THE REYNOLDS NUMBER?
%\footnote{Time integration is carried out with a 3rd-order semi-implicit algorithm~\citep{Peyret_2002,Gibson_Cvitanovic_JFM2007} with a time step of $\mathrm{\delta}t = 0.02$. Fourier spectral discretization is used in  $x$ and  $z$ with mesh spacings $\mathrm{\delta}^{+}_{x} = 8.57$ and $\mathrm{\delta}^{+}_{z} = 5.0 \sim 5.5$; Chebyshev modes are used in $y$ with $\mathrm{\delta}^{+}_{y\mathrm{, min}} = 0.081$ and $\mathrm{\delta}^{+}_{y\mathrm{, max}} = 3.7$. An artificial diffusivity term $1/(\Sc \Rey)\nabla ^2 \mbf{\alpha}$ is added to the right-hand-side of Eq.~\ref{Eq_fenep_conformation} to improve numerical stability ($\Sc = 0.5$)~\citep{Sureshkumar_Beris_POF1997}. Convergence has been checked for selected results by doubling resolution in all directions and time.}.
% Other details of the numerical method are documented elsewhere~\citep{Xi_Graham_submitted2009}.

%ADAPT THIS The streamwise box size is fixed at $L_x^+ = 360$, and the spanwise box size $L_z^+$ varies with parameters, which is determined by the method elucidated in~\citep{Xi_Graham_submitted2009}. Briefly speaking, $L_z^+$ should be the minimal size to sustain turbulence, unless it is smaller than $140$ wall units below which turbulence can only sustain near one of the walls. These box sizes are considered ``minimal'' in the sense that DNS in these MFUs only captures the essential self-sustaining structures of turbulence.

A rigorous search for MFUs would consider the parameter
          dependence of both $L_{x}^{+}$ and $L_{z}^{+}$, a task
          involving an impractically large number of simulation runs. Here we fix $L_{x}^{+} = 360$ (which is in the range of 
          streamwise sizes  of Newtonian MFU~\cite{Jimenez_Moin_JFM1991,Webber97}) and vary $L_{z}^{+}$ only. Although both length scales
          depend on parameters, $L_{z}^{+}$ is arguably the quantity of
          more interest: the dominant structures
          are the streamwise streaks and the streamwise vortices
          aligned alongside them; $L_{z}^{+}$ restricts
          the streak spacing and the size of the vortices.
%           whereas
%          $L_{x}^{+}$ only imposes a periodicity in the longitudinal
%          direction.  
%          The fact that we are able to find sustained
%          turbulence at various levels of drag reduction with fixed
%          $L_{x}^{+}$ indicates that
%          the minimal streamwise box size may not change as much as the
%          streamwise correlation length does.

% !TEX root = LiXi_hibernation.tex
%

\begin{figure}%
\centerline{\includegraphics[width = 3.25in]{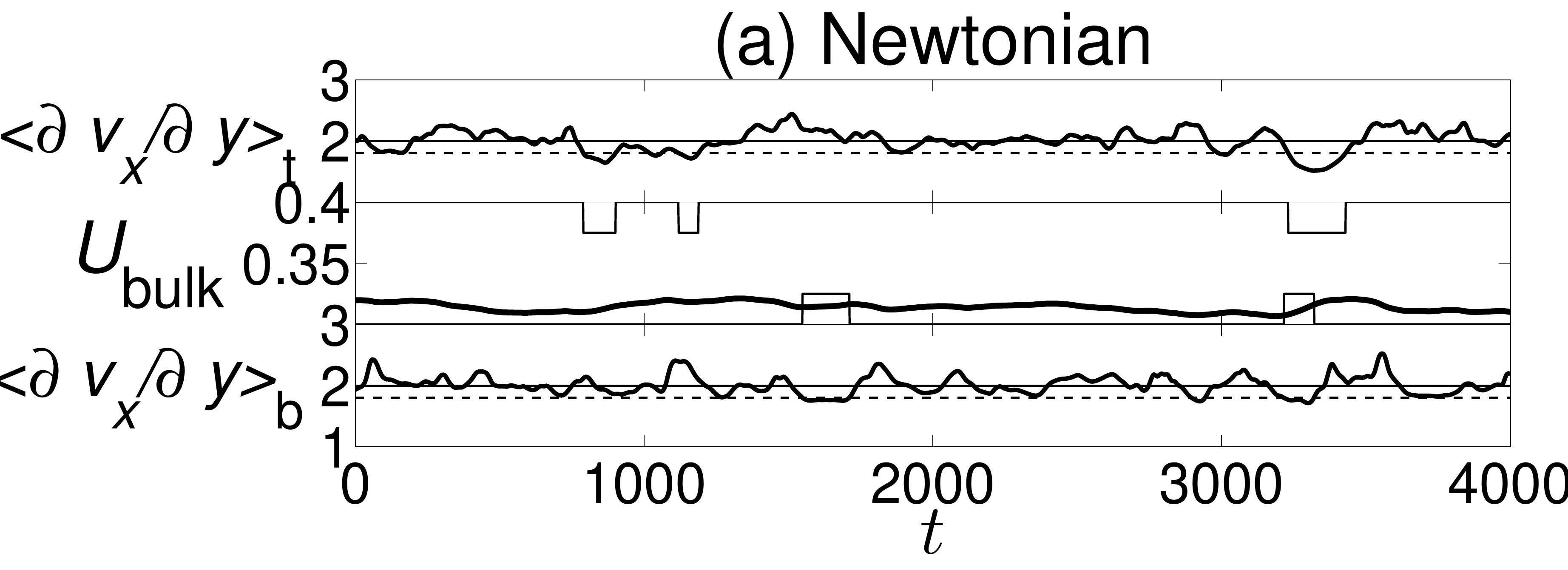}}
\centerline{\includegraphics[width = 3.25in]{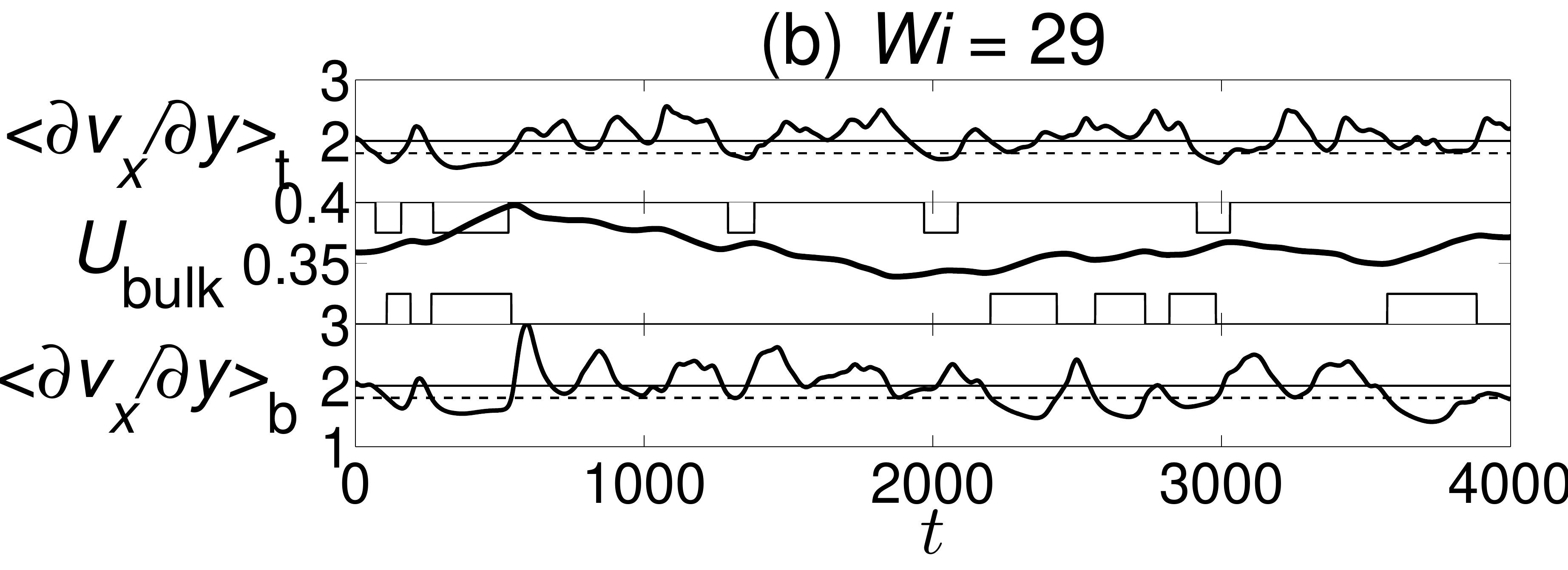}}
	\caption{Mean shear rates at the walls (``b''--bottom, ``t''--top) and bulk velocity $U_{\textrm{bulk}}$ as functions of time for typical segments of (a) Newtonian and (b) viscoelastic ($\Wi=29$) simulations.
%	 Rectangular bumps in middle panels indicate the hibernating periods at the wall of the corresponding side. 
	 Solid and dashed lines show $\langle\partial v_x /\partial y\rangle = 2$ and $\langle\partial v_x /\partial y\rangle = 1.8$, respectively. In laminar flow, $U_{\textrm{bulk}}=2/3$.} 
	\label{Fig_duxdy_Ubulk_t}%
\end{figure}%

	Fig.~\ref{Fig_duxdy_Ubulk_t} shows time series of instantaneous bulk average velocity $U_{\textrm{bulk}}$ and area-averaged shear rate $\langle \partial  v_x/\partial y\rangle $ at the top and bottom walls for (a) Newtonian flow and (b) viscoelastic flow at $\Wi=29$, where $DR\%=26$ and $L_{z}^+$ has increased from $140$ to $250$. In the Newtonian case, Fig.~\ref{Fig_duxdy_Ubulk_t}(a), one occasionally observes long-lasting periods when the shear rate at one or both walls is substantially lower than the average value of 2 -- for example the time interval $3200 < t < 3400$. By momentum conservation, the bulk velocity increases during these periods. A similar observation was made in the Newtonian MFU study of Webber \emph{et al.} \cite{Webber97}. These periods will be termed ``hibernation'', in contrast to the ``active'' turbulence found outside them.  As $\Wi$ increases, hibernation periods become increasingly frequent (Fig.~\ref{Fig_duxdy_Ubulk_t}(b)) -- since the bulk velocity increases during these periods, they contribute substantially to drag reduction. 
%	Note that the flow does not closely approach the laminar state ($U_{\textrm{bulk}}=2/3$) during hibernation periods.

To systematically identify hibernation events, two criteria are used: (1) area-averaged wall shear rate at one or both walls drops below a cutoff value $\langle\partial v_x /\partial y\rangle \vert _\mathrm{cutoff}=1.8$; and (2) it stays there for longer than a certain amount of time $\Delta t_\mathrm{cutoff}=50$. 
%Here we choose $\langle\partial v_x /\partial y\rangle \vert _\mathrm{cutoff} = 1.80$  and $\Delta t_\mathrm{cutoff} = 50$. 
%The latter is chosen so that transitory drops in $\langle \partial v_x /\partial y\rangle $ during normal turbulent fluctuations (which occur on an $O(1)$ time scale) in active periods are excluded. 
Hibernating periods so identified are shown in the middle panels of Fig.~\ref{Fig_duxdy_Ubulk_t} as rectangular signals, on the top or bottom of the plot according to the wall(s) on which the criterion is satisfied. 
%We have found that  changing the cutoff and duration values within a reasonable range does not qualitatively affect the following discussion.

\begin{figure}%
\centerline{\includegraphics[height = 1.5in]{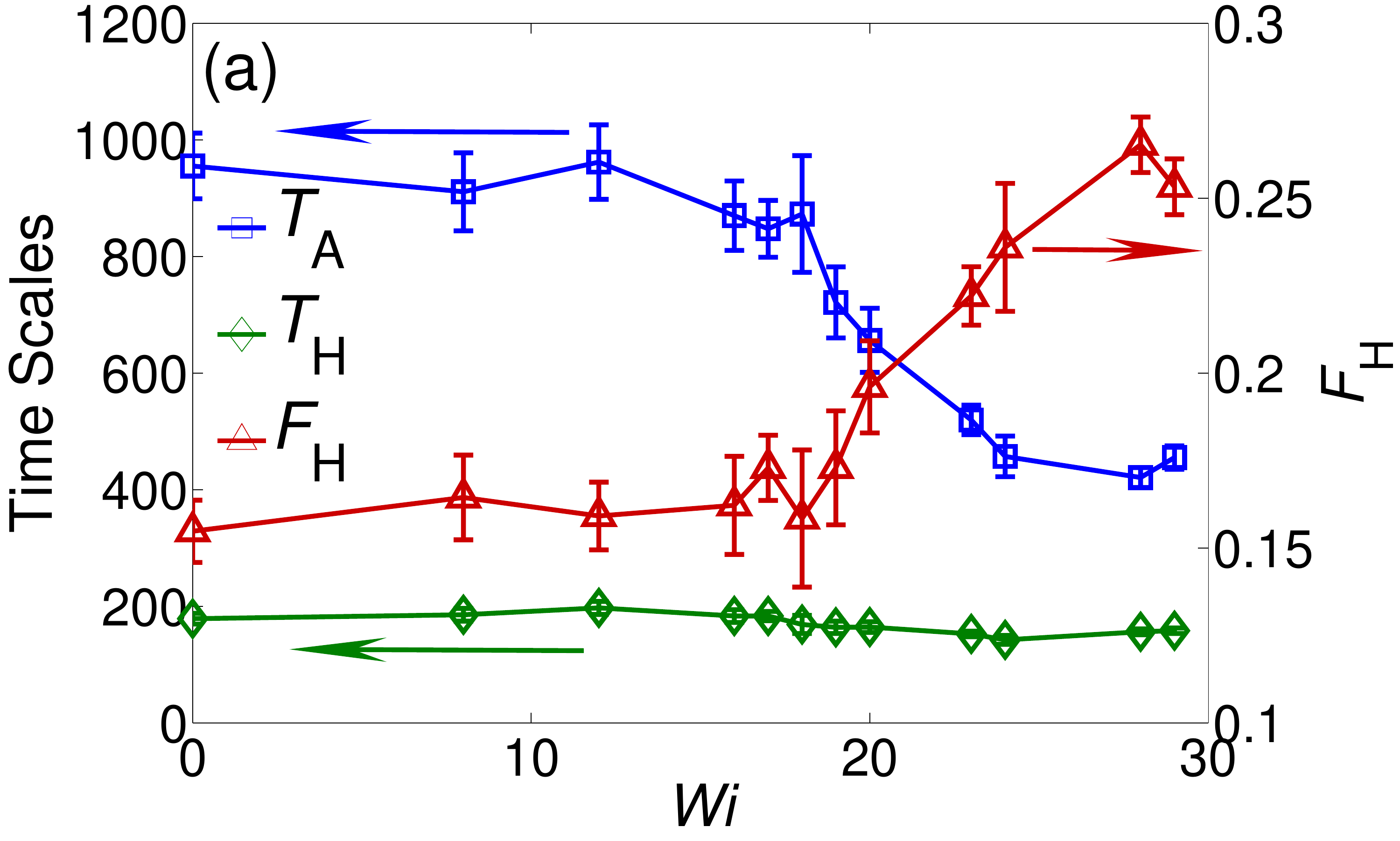}}%
\centerline{\includegraphics[height = 1.5in]{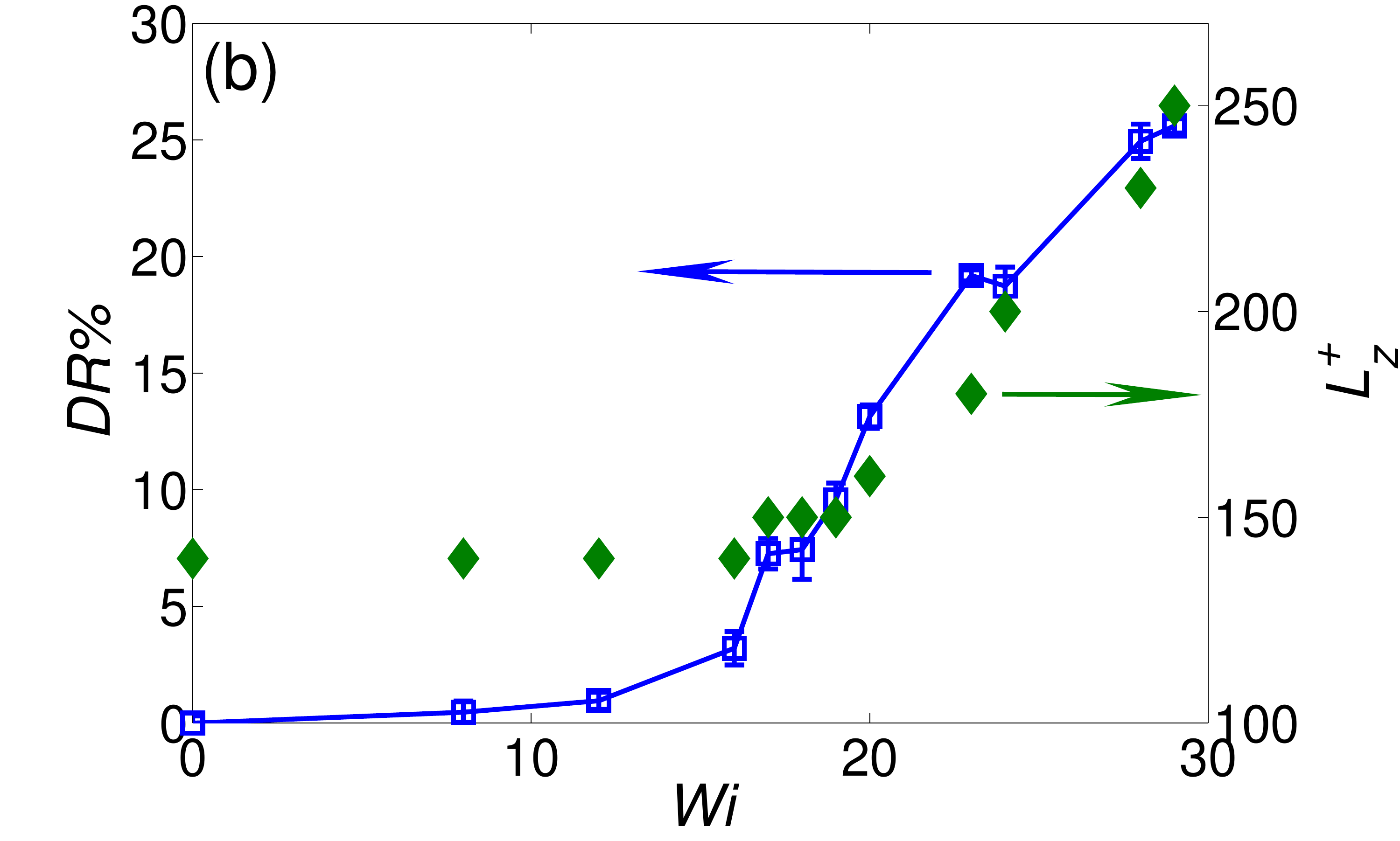}}
%\centerline{(a)\includegraphics[width =2.25in]{figures/DR_Lzp_Wi}}
%\centerline{(b)\includegraphics[width =2.25in]{figures/tscales_fraction_Wi}}%
%\centerline{(a)\includegraphics[width =1.5in]{figures/DR_Lzp_Wi}(b)\includegraphics[width =1.5in]{figures/tscales_fraction_Wi}}
 \caption{(a) Time scales and fraction of time spent in hibernation and (b) level of drag reduction and spanwise box size, as functions of $\Wi$. (At the relatively low Reynolds number considered here, the flow laminarizes for $\Wi \gtrsim 31$).}
 
%   All statistics are from statistically stationary results obtained over at least $3.2\cdot 10^{4}$ time units. }
%  $L_z^{+}$ varies with $\Wi$ and is documented in~\cite{Xi_Graham_submitted2009} -- THIS IS NOT SATISFACTORY -- CAN WE INCLUDE $L_{x}^{+}$?. Onset of drag reduction occurs at $\Wi_\mathrm{onset} \approx 16$ (WE SHOULD INCLUDE THIS ON THE PLOT), and LDR--HDR transition occurs at $\Wi_\mathrm{LDR-HDR} \approx 19$ at these parameters. For each $\Wi$, averages are taken with multiple independent simulation runs, each of which lasts for at least $8000$ time-units (TUs) excluding the initial transitional phase; the total amount of TUs included in each average is in the range of $32000 - 117700$ ($O(10\Rey)$) or longer). Error bars for time scales are estimated assuming that occurrences of hibernation are independent of one another; error bars for the percentage (FRACTION?) are estimated using block-averaging~\cite{Flyvbjerg_Petersen_JCP1989} with the block-size of $4000$ TUs.
%  }
    \label{Fig_tscales_percentage_Wi}%
\end{figure}%

%(NEED TO ADDRESS THE LOWER LIMIT OF 140 ON LZ+.)
Based on this identification scheme, Fig.~\ref{Fig_tscales_percentage_Wi}(a) shows, as functions of $\Wi$, the mean duration of the hibernation periods $T_{H}$, mean duration of active periods $T_{A}$, and fraction of time spent in hibernation $F_{H}$. Corresponding results for minimal spanwise box size and level of drag reduction are in Fig.~\ref{Fig_tscales_percentage_Wi}(b). Notice that the average duration $T_{H}$ of a hibernating period is almost completely insensitive to $\Wi$. In contrast, the average duration $T_{A}$ of an active turbulence phase decreases substantially after onset of drag reduction. 
%Accordingly, the fraction of time spent in hibernation is determined only by $T_{A}$, since $T_{H}$ does not depend on $\Wi$. 
Therefore, at high $\Wi$, viscoelasticity compresses the lifetime of active turbulence intervals, while having virtually no effect on hibernation.
% As a net outcome of these two time scales, the percentage of time that turbulence spends in hibernation remains approximately unchanged until the LDR--HDR transition, after which it starts to ascend rapidly. Therefore significant effects of increased hibernation frequency on the trend of drag reduction and flow statistics should only be expected after the LDR--HDR transition, which is consistent with the observation in both experiments~\cite{Warholic_Hanratty_EXPFL1999} and earlier work on MFU~\cite{Xi_Graham_submitted2009} that qualitative changes in turbulence statistics occur at the LDR--HDR transition. (AGAIN, HAVE WE ADEQUATELY EXPLAINED ABOUT LDR-HDR?)

\begin{figure}%
\centerline{\includegraphics[height = 1.5in]{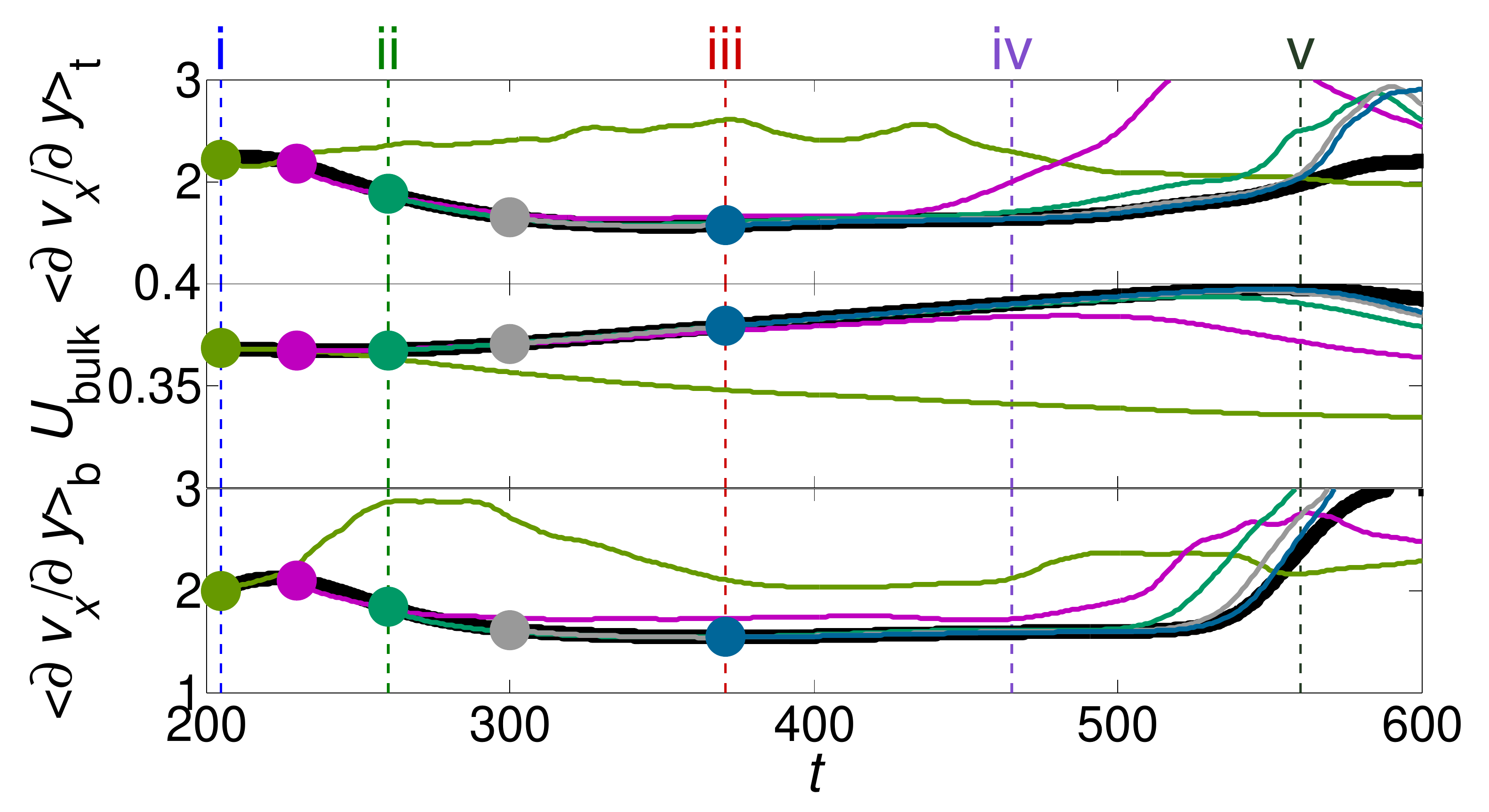}}%
    \caption{A hibernation event. Thick black lines are results at $\Wi = 29$ ($200 \leqslant t \leqslant 600$ in Fig.~\ref{Fig_duxdy_Ubulk_t}(b)). Colored lines are from Newtonian simulations started at the dots, using velocity fields from the $\Wi=29$ simulation. }
%    Labeled vertical grid lines indicate the snapshots illustrated in Figs.~\ref{fig:instantprofiles}.}
    \label{Fig_duxdy_Ubulk_t_blowup}%
\end{figure}%

The insensitivity of $T_{H}$ to $\Wi$ suggests that flow during hibernation does not strongly stretch polymer molecules. Indeed, at $\Wi=29$ the peak value of $\langle \alpha_{yy}\rangle$, which is closely associated with streamwise vortex suppression~\citep{Procaccia_Lvov_RMP2008, Stone_Graham_POF2004,Li_Graham_POF2007}, drops from about 210 in active turbulence to about 5 during hibernation, a 40-fold reduction. These results suggest that hibernation should be very similar in the Newtonian and viscoelastic cases. To test this possibility, velocity fields from time instants before and during a hibernation event at $\Wi=29$ were used as initial conditions for a Newtonian simulation, the trajectories of which were then compared with those from the original viscoelastic simulation. Fig.~\ref{Fig_duxdy_Ubulk_t_blowup} illustrates the original viscoelastic trajectory (thick black line) as well as Newtonian trajectories (colors) started at various times.
For the Newtonian run starting before any sign of hibernation is observed ($t = 205$), active turbulence is sustained. However, runs started from later times show that once the system begins to enter hibernation, removing the polymer stress does not cause turbulence to revert to an active state, although the depth and duration of hibernation are weakly dependent on the start time.
% In short, while polymer increases the probability of entering hibernation, it has little effect on flow within the hibernation region itself.
\begin{figure}
(a) ~~~~~~~~~~~~~~~~~~~~~~~~~~~~~~~~~~~~~~~~~~~~~~~~~~~~~~~~~~~~~~~~~~~~~~~~~~\\
\vspace{-0.2in}
	\centerline{\includegraphics[height = 1.5in]{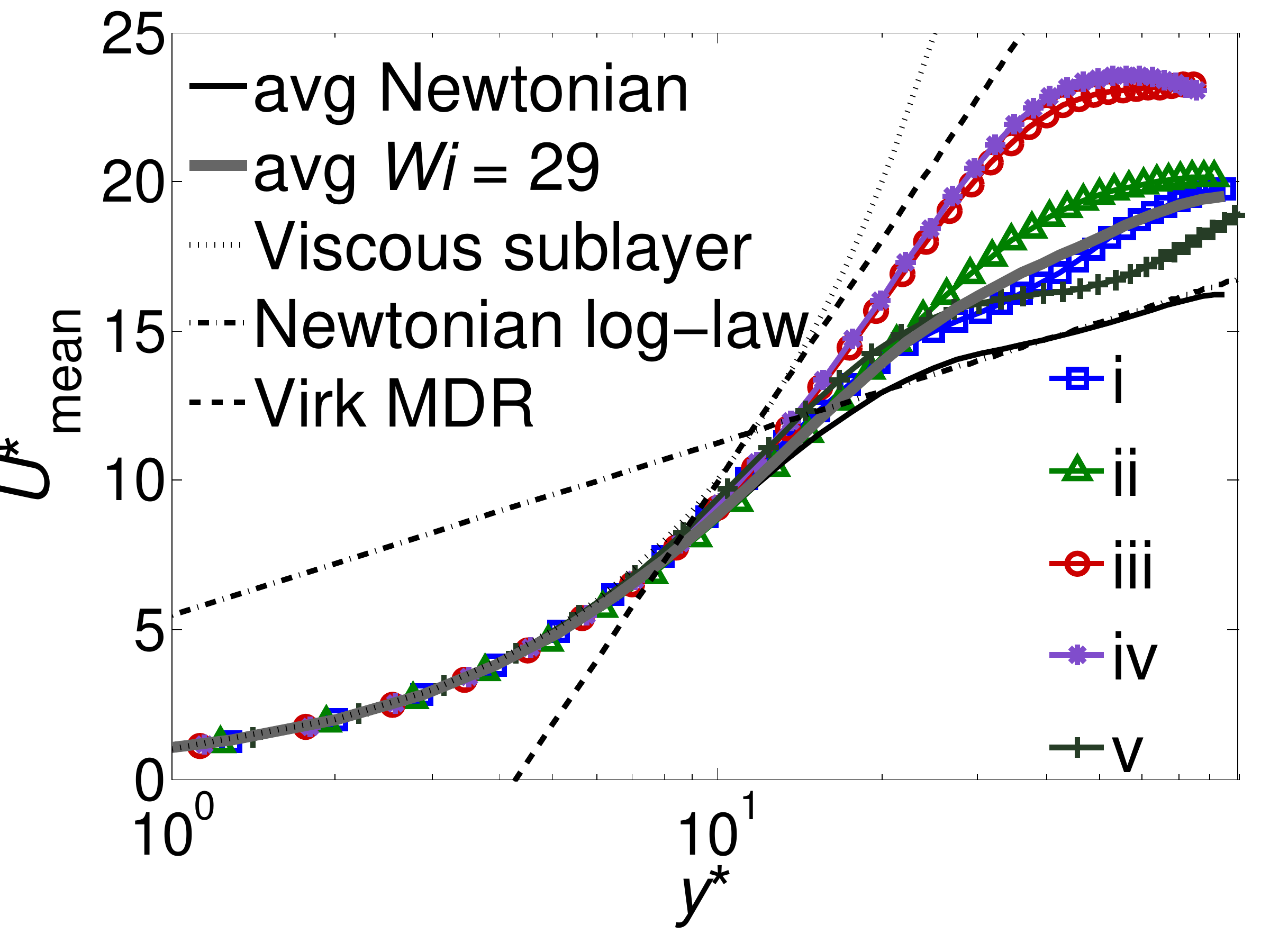}}%
(b) ~~~~~~~~~~~~~~~~~~~~~~~~~~~~~~~~~~~~~~~~~~~~~~~~~~~~~~~~~~~~~~~~~~~~~~~~~~\\
\vspace{-0.2in}
	\centerline{\includegraphics[height=1.5in]{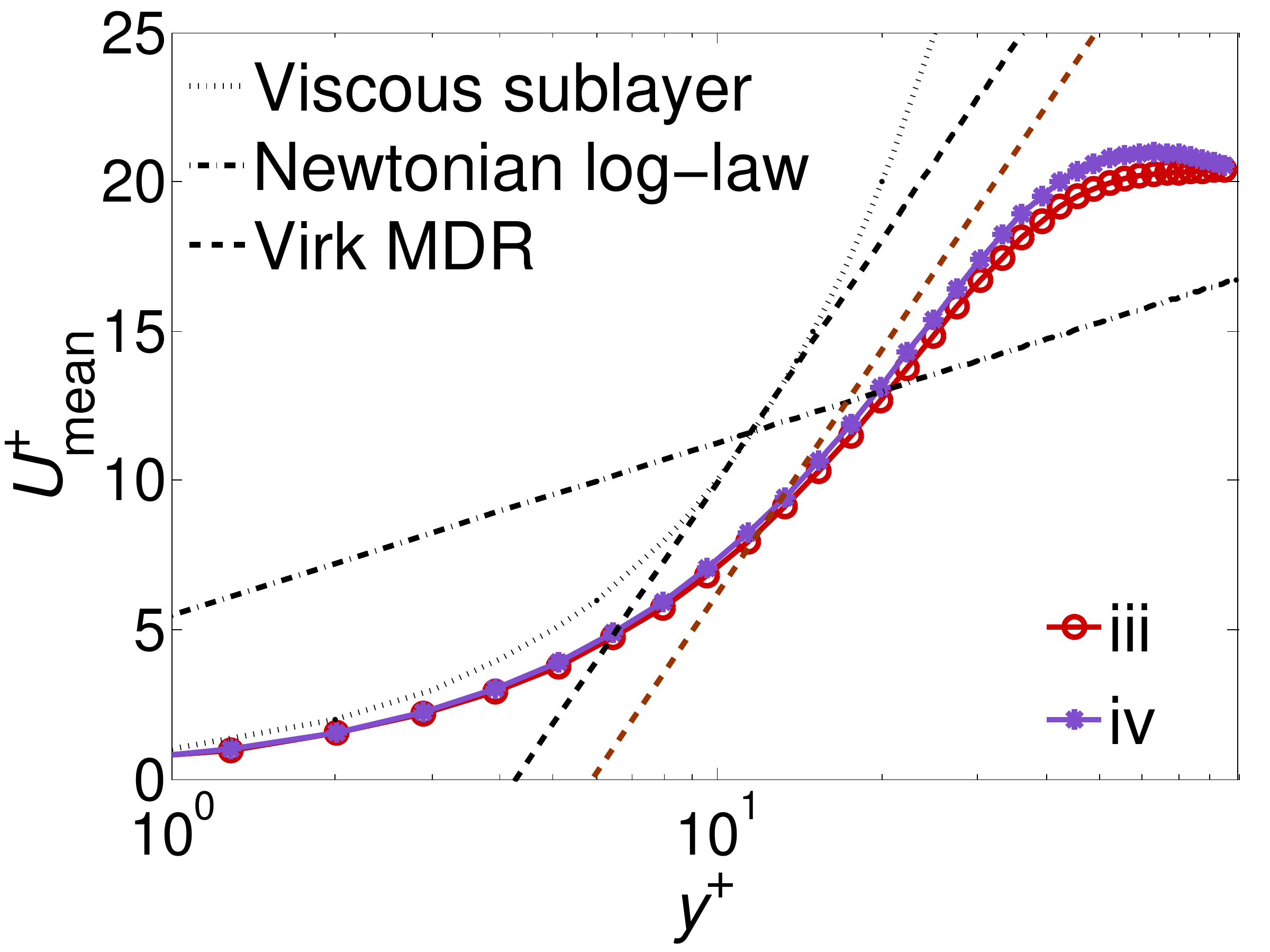}}
(c) ~~~~~~~~~~~~~~~~~~~~~~~~~~~~~~~~~~~~~~~~~~~~~~~~~~~~~~~~~~~~~~~~~~~~~~~~~~\\
\vspace{-0.2in}
	\centerline{\includegraphics[width = 1.5in]{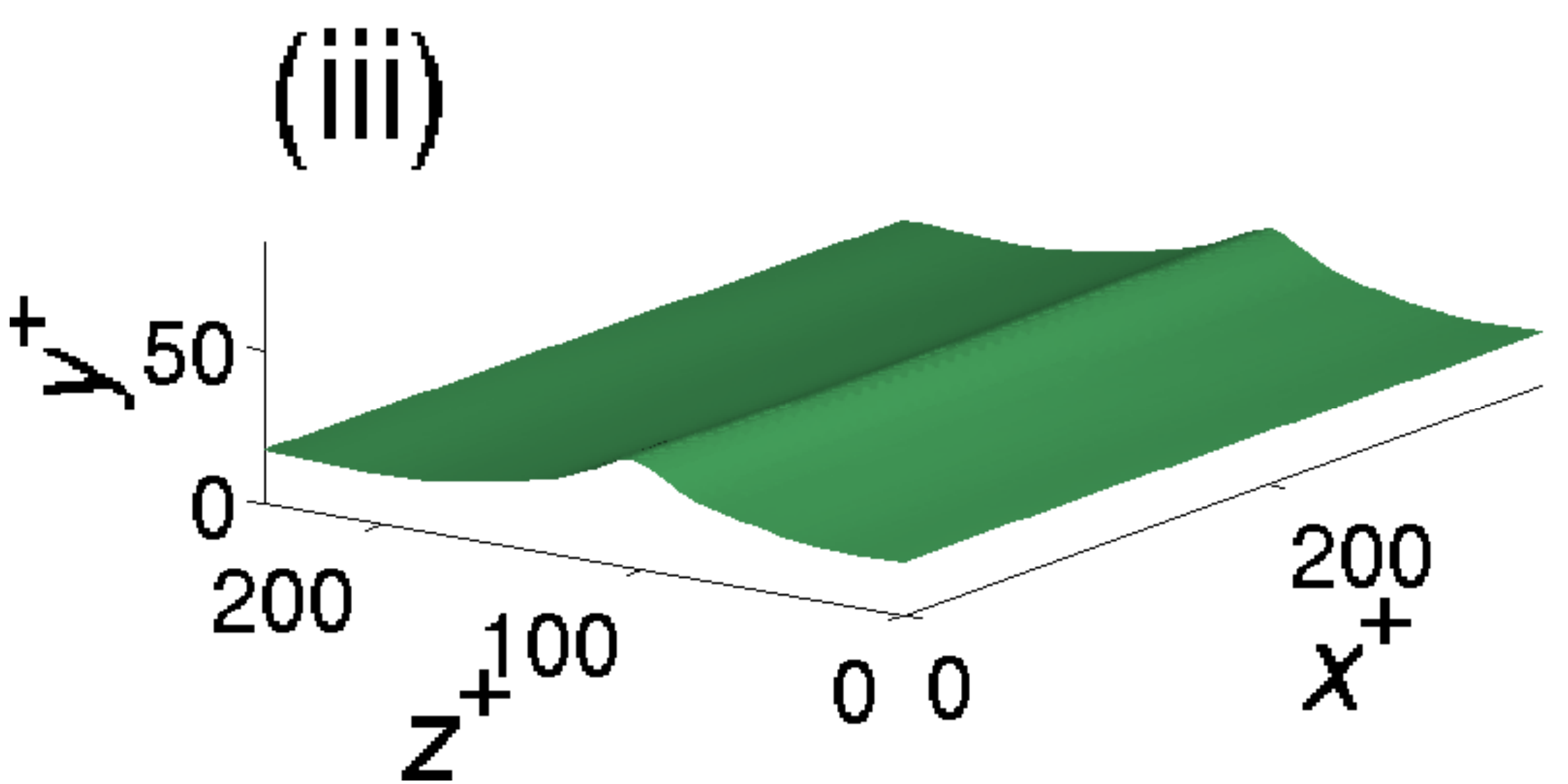}\includegraphics[width = 1.5in]{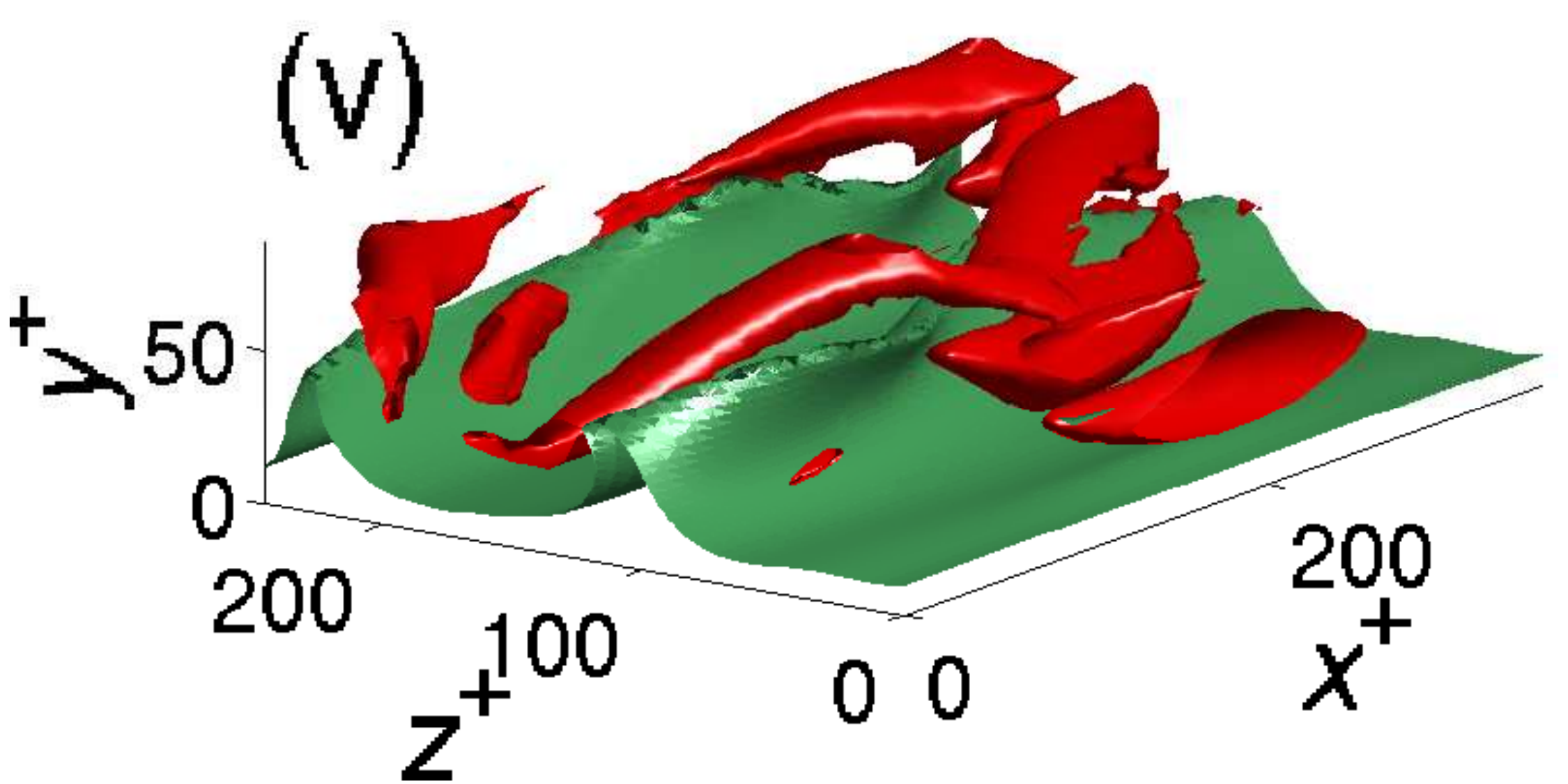}}%

%\centerline{\includegraphics[width = 3in]{figures/flowvisualization_t371}}%
%\centerline{\includegraphics[width = 3in]{figures/flowvisualization_t560}}%

%	\includegraphics[width = 0.45\textwidth]{figures/alphayy_ystar}%
%\caption{Instantaneous mean profiles (averaged in $x$ and $z$) of selected snapshots (ref. Figure~\ref{Fig_duxdy_Ubulk_t_blowup}) before, during and after a typical hibernating period. Only profiles of the bottom-half of the channel are shown. (Superscript ``*'' represents velocity and length quantities in inner-scales based on the instantaneous wall-shear stress at the wall of the corresponding side.) Top: mean velocity profiles. Black lines show typical asymptotes: ``viscous sublayer'', $U_\mathrm{mean}^* = y^*$; ``Newtonian log-law'', $U_\mathrm{mean}^* = 2.44\:\ln y^* + 5.2$~\cite{Pope_2000}; ``Virk MDR'', $U_\mathrm{mean}^* = 11.7\:\ln y^* - 17.0$~\cite{Virk_AICHEJ1975}; bottom: $\alpha_{yy}$}
\caption{(a): Instantaneous mean velocity profiles of snapshots i-v (colors), and time-averaged profiles in the Newtonian and $\Wi=29$ cases. (The latter are plotted in terms of conventional wall units.) (b): Instantaneous mean velocity profiles from time instants iii and iv plotted in conventional wall units. For comparison, a downward-shifted plot of the Virk log-law is also shown.
%Black lines show various important asymptotes: ``viscous sublayer'', $U_\mathrm{mean}^* = y^*$; ``Newtonian log-law'', $U_\mathrm{mean}^* = 2.44\:\ln y^* + 5.2$~\cite{Pope_2000}; ``Virk MDR'', $U_\mathrm{mean}^* = 11.7\:\ln y^* - 17.0$~\cite{Virk_AICHEJ1975}.
(c): flow structures of typical snapshots in hibernation (iii) and active turbulence (v). Green sheets are isosurfaces $v_x = 0.3$; pleats correspond to low-speed streaks; red tubes are isosurfaces of streamwise-vortex intensity $Q_\mathrm{2D} = 0.02$, calculated by applying the $Q$-criterion of vortex indentification~\cite{Jeong_Hussain_JFM1995} in the $yz$ plane~\cite{Li_Graham_POF2007, Xi_Graham_submitted2009}. }
% In all plots, only the bottom half of the channel is shown.}%; snapshot labels correspond to those marked in Figure~\ref{Fig_duxdy_Ubulk_t_blowup}.}
\label{fig:instantprofiles}
\end{figure}

To better understand these results, we examine more closely the hibernating period shown in Fig.~\ref{Fig_duxdy_Ubulk_t_blowup}.  Several time instants are selected as marked: (i) is just before turbulence enters hibernation; (ii) is  on the path toward hibernation; (iii) and (iv) are within hibernation; (v) is after turbulence becomes reactivated. Fig.~\ref{fig:instantprofiles}a shows instantaneous area-averaged velocity profiles in the bottom half of the channel at these instants, plotted in inner units based on the instantaneous wall shear stress at the bottom wall (denoted by the superscript $^{*}$ rather than $^{+}$). By collapsing the viscous sublayer behavior onto a single curve, this choice of scaling best exposes the nature of and differences between the various time instants shown. In active turbulence (i and v), the profiles fluctuate substantially. Profiles for snapshots completely in hibernation (iii and iv) are fundamentally different. In particular, in the range $15 \lesssim y^* \lesssim 40$, both profiles show a clear log-law relationship with a slope very close to (within $10\%$ of) the Virk MDR asymptotic slope of 11.7~\cite{Virk_AICHEJ1975}. (The differences between the intercepts is smaller than the scatter of the available data.) The figure also shows the time-averaged mean velocity profiles for the Newtonian flow (which has a log-law region in good agreement with the classical result $U_\mathrm{mean}^{+}=2.5\ln y^{+}+5.5$) and the flow at $\Wi=29$, which is, as expected, intermediate between the Newtonian and MDR profiles. 
To illustrate that the choice of scaling does not affect the conclusion that velocity profiles in hibernation display a log-law slope close to the Virk results, Fig.~\ref{fig:instantprofiles}b shows the curves from time instants iii and iv in conventional wall units. Although shifted downward from the Virk profile (because in the instantaneous wall shear stress is less than the time-averaged wall shear stress), the log-law slope remains within $20\%$ of the Virk value. The Newtonian hibernation periods (not shown, for brevity) are very similar and the Virk slope is observed there as well. 

%(DO WE NEED TO SAY ANYTHING ABOUT THE SLOPE GOING BACK DOWN AT LARGE Y*?)  WE NEED TO ALSO ADDRESS THE ISSUE OF THE REYNOLDS SHEAR STRESSES -- POINT OUT THAT THEY BECOME VERY SMALL.
%On the orbit toward hibernation (b), the profile is qualitatively similar expect that the slope is lower than that of Virk MDR.

Fig.~\ref{fig:instantprofiles}c shows flow structures corresponding to time instants iii and v. Within active periods (v), turbulence shows the expected highly 3D structure of streamwise vortices and low-speed streaks~\cite{Robinson_ARFM1991, Jimenez_Moin_JFM1991, Waleffe_POF1997}.
%, although for the high $\Wi$ run selected here, very often two streaks are involved in the MFU~\cite{Xi_Graham_submitted2009}.
 During hibernation (iii), streamwise vortices are significantly weaker; low-speed streaks are still observed, but are weak and only weakly dependent on $x$. (The low shear events observed in the Newtonian MFU study of Webber \emph{et al.}  \cite{Webber97} also display weak streamwise dependence.)  Weak streamwise vorticity and three-dimensionality are distinct characteristics of the MDR regime \citep{Virk_AICHEJ1975, White_Mungal_EXPFL2004, Housiadas_Beris_POF2005, Li_Sureshkumar_JNNFM2006, White_Mungal_ARFM2008}. 
% After roughly 180 time units (the mean duration $T_{H}$ discussed earlier), the streamwise vortices and three-dimensionality grow again and the turbulence reactivates.
 The weak effect of viscoelasticity on hibernating turbulence may lie in its nearly streamwise-invariant kinematics. In the limiting case of a streamwise invariant steady flow, material lines cannot stretch exponentially \cite{OttinoBook}; accordingly, polymer stretch in such a flow will not be substantial.
 Finally, the Reynolds shear stress during hibernation drops to very low values relative to active turbulence, where it peaks at about 0.8; the peak value during hibernation is about $0.3$. Again, this result is consistent with observations in the MDR regime \cite{HulsenMDR01,Ptasinski_Nieuwstadt_JFM2003,Warholic_Hanratty_EXPFL1999, Warholic01}.  

The qualitative picture that emerges from these simulations is thus the following. Active turbulence generates substantial stretching of polymer molecules. The resulting stresses act to suppress this turbulence and drive the flow toward a very weakly turbulent hibernating regime. During hibernation the polymer molecules are no longer strongly stretched and relax toward  equilibrium.  Eventually, hibernation ends as new turbulent fluctuations begin to grow, and the system transits back into active turbulence. The active turbulence again stretches polymer chains and the (stochastic) cycle repeats.

In this picture, experimental observations in which the Virk MDR mean velocity profile is found correspond to a limiting situation -- not achieved at the low Reynolds number and small boxes studied here -- where the fraction of time and space occupied by active turbulence becomes small enough that the hibernating regime dominates the statistics. Active turbulence cannot vanish entirely, because it is known that on average, the polymer molecules carry a substantial fraction of the mean shear stress~\citep{Warholic_Hanratty_EXPFL1999, Ptasinski_Nieuwstadt_JFM2003}, and since hibernating turbulence does not stretch polymers, some active turbulence must remain. These considerations lead to a new picture of turbulence in the MDR regime as a state in which hibernating turbulence is the norm, with active turbulence arising intermittently in space and time only to be suppressed by the polymer stretching that it induces.

% In this picture, MDR is asymptotically independent of polymer properties because hibernating turbulence, which dominates the statistics of MDR, is a fundamentally Newtonian phenomenon.

%Only a study of turbulence in MFUs would allow for a picture this clean to emerge: in a larger flow domain there are likely to be some regions where the turbulence is active and some where it is hibernating, but without knowing in advance about these regions they would be difficult to identify.

%The present study focused on MFU flows at low Reynolds number, where there is not yet a large separation between inner and outer scales. This approach allowed the collection and analysis of an extensive data set  in a regime where flow structures are relatively simple. 

This study focused on MFU flows at low Reynolds number, allowing analysis of an extensive data set  in a regime where flow structures are relatively simple.
 Remarkably, even this regime displays clear signatures of the features of MDR commonly associated with higher Reynolds numbers. (Though it should be noted that the MDR asymptote is experimentally observed at values of $\Rey$ all the way down into the laminar-turbulent transition regime \cite{Virk_AICHEJ1975}.) To carefully evaluate the generality of  picture just presented, future work will require simulations at higher $\Rey$ and large boxes in combination with pattern analysis tools that can identify spatiotemporally localized regions of active and hibernating turbulence. 

%In addition, attention must be focused on the hibernating turbulence phenomenon.  Recently, Waleffe has identified a class of nonlinear traveling wave solutions to the Navier-Stokes equations in the plane Couette and Poiseuille geometries that share many characteristics with hibernating turbulence, specifically weak streamwise vortices and weak streamwise dependence \cite{Wang_Waleffe_PRL2007}. Indeed, at least one family of these solutions has vanishing streamwise dependence and Reynolds shear stress as $\Rey\rightarrow\infty$. These states are saddle points in phase space and it may be that hibernating turbulence is a trajectory moving transiently in the vicinity of one of these saddles. SOMETHING ABOUT EDGE STATES

In addition, attention must be focused on the hibernation phenomenon.  Waleffe has identified a class of nonlinear traveling wave solutions in the plane Couette and Poiseuille geometries -- saddle points in phase space -- that share many characteristics with hibernating turbulence, specifically weak streamwise vortices and weak streamwise dependence \cite{Wang_Waleffe_PRL2007}. At least one family of these solutions has vanishing streamwise dependence and Reynolds shear stress as $\Rey\rightarrow\infty$.  Similarly, other researchers \cite{Schneider_Eckhardt_PRL2007,Schneider:2008p5732,Duguet_Kerswell_JFM2008} have identified both simple and chaotic saddle trajectories, ``edge states'',  that  lie on the boundary between laminar and turbulent dynamics. These observations may allow us to make rigorous the qualitative notion described in the introduction that the MDR regime is transitional or marginal -- it may be that hibernating turbulence is very near in phase space to the boundary between laminar and turbulent flow. Finally, a better understanding of hibernation may lead to strategies for active flow control to maintain turbulence in a state of hibernation and thus dramatically reduce skin friction.

%A problem that remains open in this picture is then the self-sustaining mechanism that keeps hibernating turbulence in existence. A potential solution to this problem lies in the lower branch coherent states discussed above. Recall that Waleffe has shown that one family of these solutions has asymptotically finite amplitude but vanishing streamwise dependence as $Re\rightarrow\infty$ \cite{Wang_Waleffe_PRL2007}. These states are saddle points and it may be that hibernating turbulence is simply a phase space trajectory moving in the vicinity of one of these states.

%These hypotheses, that MDR turbulence is fundamentally hibernating turbulence, and that hibernating turbulence is closely related to nonlinear traveling wave structures in Newtonian flow, point toward a fundamentally new direction for research in the field of turbulent drag reduction by additives, which will be the topic of future work.

% !TEX root = LiXi_hibernation.tex
%
The authors thank Fabian~Waleffe for many helpful discussions. The code used here is based on \textit{ChannelFlow} by John F.~Gibson, whose assistance we gratefully acknowledge. This work is supported by the National Science Foundation (DDDAS-SMRP-0540147, CBET-0730006).

%\bibliography{drprop}%,other}

\end{document}